\documentclass[lettersize]{IEEEtran}
\usepackage{graphicx}
\usepackage{subfigure}
\usepackage{multirow}
\usepackage{array}
\usepackage{caption}
\newcolumntype{P}[1]{>{\centering\arraybackslash}p{#1}}
\newcolumntype{M}[1]{>{\centering\arraybackslash}m{#1}}
\usepackage{graphicx}
\usepackage{subfigure}

\usepackage{amsthm,amssymb,amsmath,multirow,color,amsfonts}%
\usepackage[update,prepend]{epstopdf}
\usepackage{multirow}
\usepackage[latin1]{inputenc}
\usepackage{tikz}
\usepackage{bbm} 
\usepackage{pdfpages}
\usepackage{multirow}
\usepackage{subfig}
\usepackage{comment}
\usepackage{multicol}
\usepackage{cite}

\usepackage[justification=centering]{caption}
\usepackage{textcomp}
\usepackage{psfrag}
\usepackage{arydshln}
\usepackage{url}
\usepackage{soul}
\usepackage{graphicx,color}
\usepackage[nolist]{acronym}
\usepackage{algorithm,algorithmic} 

\usepackage{mathtools,lipsum}
\usepackage{cuted}
\usepackage{amsmath}
 
\usepackage{mathrsfs}
\usepackage{color}

\usepackage[capitalise]{cleveref}
\Crefname{equation}{Eq.\!}{Eqs.\!}
\Crefname{figure}{Fig.\!}{Figs.\!}
\Crefname{tabular}{Tab.\!}{Tabs.\!}
\Crefname{section}{Section\!}{Sections.\!}

\usepackage{graphicx}
\graphicspath{ {images/} }

\newenvironment{sequation}{
 \begin{equation}\small } { \end{equation}
}

\def\nb0{{\mathbf{0}}}
\def\nb1{{\mathbf{1}}}

\newtheorem{lemma}{Lemma}

\newtheorem{theorem}{Theorem}

\def\R{\mathbb{R}}

\begin{document}
\graphicspath{{./Figures/}}
\begin{acronym}

\acro{RF}{radio frequency}
\acro{THz}{terahertz}
\acro{HRTRN}{hybrid RF and THz relay network}

\acro{5G-NR}{5G New Radio}
\acro{3GPP}{3rd Generation Partnership Project}
\acro{ABS}{aerial base station}
\acro{AC}{address coding}
\acro{ACF}{autocorrelation function}
\acro{ACR}{autocorrelation receiver}
\acro{ADC}{analog-to-digital converter}
\acrodef{aic}[AIC]{Analog-to-Information Converter}     
\acro{AIC}[AIC]{Akaike information criterion}
\acro{aric}[ARIC]{asymmetric restricted isometry constant}
\acro{arip}[ARIP]{asymmetric restricted isometry property}

\acro{ARQ}{Automatic Repeat Request}
\acro{AUB}{asymptotic union bound}
\acrodef{awgn}[AWGN]{Additive White Gaussian Noise}     
\acro{AWGN}{additive white Gaussian noise}

\acro{APSK}[PSK]{asymmetric PSK} 

\acro{waric}[AWRICs]{asymmetric weak restricted isometry constants}
\acro{warip}[AWRIP]{asymmetric weak restricted isometry property}
\acro{BCH}{Bose, Chaudhuri, and Hocquenghem}        
\acro{BCHC}[BCHSC]{BCH based source coding}
\acro{BEP}{bit error probability}
\acro{BFC}{block fading channel}
\acro{BG}[BG]{Bernoulli-Gaussian}
\acro{BGG}{Bernoulli-Generalized Gaussian}
\acro{BPAM}{binary pulse amplitude modulation}
\acro{BPDN}{Basis Pursuit Denoising}
\acro{BPPM}{binary pulse position modulation}
\acro{BPSK}{Binary Phase Shift Keying}
\acro{BPZF}{bandpass zonal filter}
\acro{BSC}{binary symmetric channels}              
\acro{BU}[BU]{Bernoulli-uniform}
\acro{BER}{bit error rate}
\acro{BS}{base station}
\acro{BW}{BandWidth}
\acro{BLLL}{ binary log-linear learning }

\acro{CP}{Cyclic Prefix}
\acrodef{cdf}[CDF]{cumulative distribution function}   
\acro{CDF}{Cumulative Distribution Function}
\acrodef{c.d.f.}[CDF]{cumulative distribution function}
\acro{CCDF}{complementary cumulative distribution function}
\acrodef{ccdf}[CCDF]{complementary CDF}               
\acrodef{c.c.d.f.}[CCDF]{complementary cumulative distribution function}
\acro{CD}{cooperative diversity}

\acro{CDMA}{Code Division Multiple Access}
\acro{ch.f.}{characteristic function}
\acro{CIR}{channel impulse response}
\acro{cosamp}[CoSaMP]{compressive sampling matching pursuit}
\acro{CR}{cognitive radio}
\acro{cs}[CS]{compressed sensing}                   
\acrodef{cscapital}[CS]{Compressed sensing} 
\acrodef{CS}[CS]{compressed sensing}
\acro{CSI}{channel state information}
\acro{CCSDS}{consultative committee for space data systems}
\acro{CC}{convolutional coding}
\acro{Covid19}[COVID-19]{Coronavirus disease}

\acro{DAA}{detect and avoid}
\acro{DAB}{digital audio broadcasting}
\acro{DCT}{discrete cosine transform}
\acro{dft}[DFT]{discrete Fourier transform}
\acro{DR}{distortion-rate}
\acro{DS}{direct sequence}
\acro{DS-SS}{direct-sequence spread-spectrum}
\acro{DTR}{differential transmitted-reference}
\acro{DVB-H}{digital video broadcasting\,--\,handheld}
\acro{DVB-T}{digital video broadcasting\,--\,terrestrial}
\acro{DL}{DownLink}
\acro{DSSS}{Direct Sequence Spread Spectrum}
\acro{DFT-s-OFDM}{Discrete Fourier Transform-spread-Orthogonal Frequency Division Multiplexing}
\acro{DAS}{Distributed Antenna System}
\acro{DNA}{DeoxyriboNucleic Acid}

\acro{EC}{European Commission}
\acro{EED}[EED]{exact eigenvalues distribution}
\acro{EIRP}{Equivalent Isotropically Radiated Power}
\acro{ELP}{equivalent low-pass}
\acro{eMBB}{Enhanced Mobile Broadband}
\acro{EMF}{ElectroMagnetic Field}
\acro{EU}{European union}
\acro{EI}{Exposure Index}
\acro{eICIC}{enhanced Inter-Cell Interference Coordination}

\acro{FC}[FC]{fusion center}
\acro{FCC}{Federal Communications Commission}
\acro{FEC}{forward error correction}
\acro{FFT}{fast Fourier transform}
\acro{FH}{frequency-hopping}
\acro{FH-SS}{frequency-hopping spread-spectrum}
\acrodef{FS}{Frame synchronization}
\acro{FSsmall}[FS]{frame synchronization}  
\acro{FDMA}{Frequency Division Multiple Access}

\acro{GA}{Gaussian approximation}
\acro{GF}{Galois field }
\acro{GG}{Generalized-Gaussian}
\acro{GIC}[GIC]{generalized information criterion}
\acro{GLRT}{generalized likelihood ratio test}
\acro{GPS}{Global Positioning System}
\acro{GMSK}{Gaussian Minimum Shift Keying}
\acro{GSMA}{Global System for Mobile communications Association}
\acro{GS}{ground station}
\acro{GMG}{ Grid-connected MicroGeneration}

\acro{HAP}{high altitude platform}
\acro{HetNet}{Heterogeneous network}

\acro{IDR}{information distortion-rate}
\acro{IFFT}{inverse fast Fourier transform}
\acro{iht}[IHT]{iterative hard thresholding}
\acro{i.i.d.}{independent, identically distributed}
\acro{IoT}{Internet of Things}                      
\acro{IR}{impulse radio}
\acro{lric}[LRIC]{lower restricted isometry constant}
\acro{lrict}[LRICt]{lower restricted isometry constant threshold}
\acro{ISI}{intersymbol interference}
\acro{ITU}{International Telecommunication Union}
\acro{ICNIRP}{International Commission on Non-Ionizing Radiation Protection}
\acro{IEEE}{Institute of Electrical and Electronics Engineers}
\acro{ICES}{IEEE international committee on electromagnetic safety}
\acro{IEC}{International Electrotechnical Commission}
\acro{IARC}{International Agency on Research on Cancer}
\acro{IS-95}{Interim Standard 95}

\acro{KPI}{Key Performance Indicator}

\acro{LEO}{low earth orbit}
\acro{LF}{likelihood function}
\acro{LLF}{log-likelihood function}
\acro{LLR}{log-likelihood ratio}
\acro{LLRT}{log-likelihood ratio test}
\acro{LoS}{Line-of-Sight}
\acro{LRT}{likelihood ratio test}
\acro{wlric}[LWRIC]{lower weak restricted isometry constant}
\acro{wlrict}[LWRICt]{LWRIC threshold}
\acro{LPWAN}{Low Power Wide Area Network}
\acro{LoRaWAN}{Low power long Range Wide Area Network}
\acro{NLoS}{Non-Line-of-Sight}
\acro{LiFi}[Li-Fi]{light-fidelity}
 \acro{LED}{light emitting diode}
 \acro{LABS}{LoS transmission with each ABS}
 \acro{NLABS}{NLoS transmission with each ABS}

\acro{MB}{multiband}
\acro{MC}{macro cell}
\acro{MDS}{mixed distributed source}
\acro{MF}{matched filter}
\acro{m.g.f.}{moment generating function}
\acro{MI}{mutual information}
\acro{MIMO}{Multiple-Input Multiple-Output}
\acro{MISO}{multiple-input single-output}
\acrodef{maxs}[MJSO]{maximum joint support cardinality}                       
\acro{ML}[ML]{maximum likelihood}
\acro{MMSE}{minimum mean-square error}
\acro{MMV}{multiple measurement vectors}
\acrodef{MOS}{model order selection}
\acro{M-PSK}[${M}$-PSK]{$M$-ary phase shift keying}                       
\acro{M-APSK}[${M}$-PSK]{$M$-ary asymmetric PSK} 
\acro{MP}{ multi-period}
\acro{MINLP}{mixed integer non-linear programming}

\acro{M-QAM}[$M$-QAM]{$M$-ary quadrature amplitude modulation}
\acro{MRC}{maximal ratio combiner}                  
\acro{maxs}[MSO]{maximum sparsity order}                                      
\acro{M2M}{Machine-to-Machine}                                                
\acro{MUI}{multi-user interference}
\acro{mMTC}{massive Machine Type Communications}      
\acro{mm-Wave}{millimeter-wave}
\acro{MP}{mobile phone}
\acro{MPE}{maximum permissible exposure}
\acro{MAC}{media access control}
\acro{NB}{narrowband}
\acro{NBI}{narrowband interference}
\acro{NLA}{nonlinear sparse approximation}
\acro{NLOS}{Non-Line of Sight}
\acro{NTIA}{National Telecommunications and Information Administration}
\acro{NTP}{National Toxicology Program}
\acro{NHS}{National Health Service}

\acro{LOS}{Line of Sight}

\acro{OC}{optimum combining}                             
\acro{OC}{optimum combining}
\acro{ODE}{operational distortion-energy}
\acro{ODR}{operational distortion-rate}
\acro{OFDM}{Orthogonal Frequency-Division Multiplexing}
\acro{omp}[OMP]{orthogonal matching pursuit}
\acro{OSMP}[OSMP]{orthogonal subspace matching pursuit}
\acro{OQAM}{offset quadrature amplitude modulation}
\acro{OQPSK}{offset QPSK}
\acro{OFDMA}{Orthogonal Frequency-division Multiple Access}
\acro{OPEX}{Operating Expenditures}
\acro{OQPSK/PM}{OQPSK with phase modulation}

\acro{PAM}{pulse amplitude modulation}
\acro{PAR}{peak-to-average ratio}
\acrodef{pdf}[PDF]{probability density function}                      
\acro{PDF}{probability density function}
\acrodef{p.d.f.}[PDF]{probability distribution function}
\acro{PDP}{power dispersion profile}
\acro{PMF}{probability mass function}                             
\acrodef{p.m.f.}[PMF]{probability mass function}
\acro{PN}{pseudo-noise}
\acro{PPM}{pulse position modulation}
\acro{PRake}{Partial Rake}
\acro{PSD}{power spectral density}
\acro{PSEP}{pairwise synchronization error probability}
\acro{PSK}{phase shift keying}
\acro{PD}{power density}
\acro{8-PSK}[$8$-PSK]{$8$-phase shift keying}
\acro{PPP}{Poisson point process}
\acro{PCP}{Poisson cluster process}
 
\acro{FSK}{Frequency Shift Keying}

\acro{QAM}{Quadrature Amplitude Modulation}
\acro{QPSK}{Quadrature Phase Shift Keying}
\acro{OQPSK/PM}{OQPSK with phase modulator }

\acro{RD}[RD]{raw data}
\acro{RDL}{"random data limit"}
\acro{ric}[RIC]{restricted isometry constant}
\acro{rict}[RICt]{restricted isometry constant threshold}
\acro{rip}[RIP]{restricted isometry property}
\acro{ROC}{receiver operating characteristic}
\acro{rq}[RQ]{Raleigh quotient}
\acro{RS}[RS]{Reed-Solomon}
\acro{RSC}[RSSC]{RS based source coding}
\acro{r.v.}{random variable}                               
\acro{R.V.}{random vector}
\acro{RMS}{root mean square}
\acro{RFR}{radiofrequency radiation}
\acro{RIS}{Reconfigurable Intelligent Surface}
\acro{RNA}{RiboNucleic Acid}
\acro{RRM}{Radio Resource Management}
\acro{RUE}{reference user equipments}
\acro{RAT}{radio access technology}
\acro{RB}{resource block}

\acro{SA}[SA-Music]{subspace-augmented MUSIC with OSMP}
\acro{SC}{small cell}
\acro{SCBSES}[SCBSES]{Source Compression Based Syndrome Encoding Scheme}
\acro{SCM}{sample covariance matrix}
\acro{SEP}{symbol error probability}
\acro{SG}[SG]{sparse-land Gaussian model}
\acro{SIMO}{single-input multiple-output}
\acro{SINR}{signal-to-interference plus noise ratio}
\acro{SIR}{signal-to-interference ratio}
\acro{SISO}{Single-Input Single-Output}
\acro{SMV}{single measurement vector}
\acro{SNR}[\textrm{SNR}]{signal-to-noise ratio} 
\acro{sp}[SP]{subspace pursuit}
\acro{SS}{spread spectrum}
\acro{SW}{sync word}
\acro{SAR}{specific absorption rate}
\acro{SSB}{synchronization signal block}
\acro{SR}{shrink and realign}

\acro{tUAV}{tethered Unmanned Aerial Vehicle}
\acro{TBS}{terrestrial base station}

\acro{uUAV}{untethered Unmanned Aerial Vehicle}
\acro{PDF}{probability density functions}

\acro{PL}{path-loss}

\acro{TH}{time-hopping}
\acro{ToA}{time-of-arrival}
\acro{TR}{transmitted-reference}
\acro{TW}{Tracy-Widom}
\acro{TWDT}{TW Distribution Tail}
\acro{TCM}{trellis coded modulation}
\acro{TDD}{Time-Division Duplexing}
\acro{TDMA}{Time Division Multiple Access}
\acro{Tx}{average transmit}

\acro{UAV}{Unmanned Aerial Vehicle}
\acro{uric}[URIC]{upper restricted isometry constant}
\acro{urict}[URICt]{upper restricted isometry constant threshold}
\acro{UWB}{ultrawide band}
\acro{UWBcap}[UWB]{Ultrawide band}   
\acro{URLLC}{Ultra Reliable Low Latency Communications}
         
\acro{wuric}[UWRIC]{upper weak restricted isometry constant}
\acro{wurict}[UWRICt]{UWRIC threshold}                
\acro{UE}{User Equipment}
\acro{UL}{UpLink}

\acro{WiM}[WiM]{weigh-in-motion}
\acro{WLAN}{wireless local area network}
\acro{wm}[WM]{Wishart matrix}                               
\acroplural{wm}[WM]{Wishart matrices}
\acro{WMAN}{wireless metropolitan area network}
\acro{WPAN}{wireless personal area network}
\acro{wric}[WRIC]{weak restricted isometry constant}
\acro{wrict}[WRICt]{weak restricted isometry constant thresholds}
\acro{wrip}[WRIP]{weak restricted isometry property}
\acro{WSN}{wireless sensor network}                        
\acro{WSS}{Wide-Sense Stationary}
\acro{WHO}{World Health Organization}
\acro{Wi-Fi}{Wireless Fidelity}

\acro{sss}[SpaSoSEnc]{sparse source syndrome encoding}

\acro{VLC}{Visible Light Communication}
\acro{VPN}{Virtual Private Network} 
\acro{RF}{Radio Frequency}
\acro{FSO}{Free Space Optics}
\acro{IoST}{Internet of Space Things}

\acro{GSM}{Global System for Mobile Communications}
\acro{2G}{Second-generation cellular network}
\acro{3G}{Third-generation cellular network}
\acro{4G}{Fourth-generation cellular network}
\acro{5G}{Fifth-generation cellular network}	
\acro{gNB}{next-generation Node-B Base Station}
\acro{NR}{New Radio}
\acro{UMTS}{Universal Mobile Telecommunications Service}
\acro{LTE}{Long Term Evolution}

\acro{QoS}{Quality of Service}
\end{acronym}
	
\newcommand{\SAR} {\mathrm{SAR}}
\newcommand{\WBSAR} {\mathrm{SAR}_{\mathsf{WB}}}
\newcommand{\gSAR} {\mathrm{SAR}_{10\si{\gram}}}
\newcommand{\Sab} {S_{\mathsf{ab}}}
\newcommand{\Eavg} {E_{\mathsf{avg}}}
\newcommand{\ft}{f_{\textsf{th}}}
\newcommand{\alphatf}{\alpha_{24}}

\title{
\vspace{-11mm}
Coverage Analysis of Hybrid RF/THz Networks With
Best Relay Selection
}
\vspace{-4mm}
\author{
Zhengying Lou, Baha Eddine Youcef Belmekki, and Mohamed-Slim Alouini, {\em Fellow, IEEE}
\thanks{The authors are with King Abdullah University of Science and Technology (KAUST), CEMSE division, Thuwal 23955-6900, Saudi Arabia (e-mail: zhengying.lou@kaust.edu.sa;  bahaeddine.belmekki@kaust.edu.sa;  slim.alouini@kaust.edu.sa).}
\vspace{-12mm}
}
\maketitle

\begin{abstract}
Utilizing terahertz (THz) transmission to enhance coverage has proven various benefits compared to traditional radio frequency (RF) counterparts. This letter proposes a dual-hop decode-and-forward (DF) routing protocol in a hybrid RF and THz relay network named hybrid relay selection (HRS). The coverage probability of the HRS protocol is derived. The HRS protocol prioritizes THz relays for higher data rates or short source-destination distances; and RF relays for lower data rates or large source-destination distances. The proposed HRS protocol offers nearly the same performance as the optimal selection protocol, which requires complete instantaneous channel state information (CSI) of all the nodes. 

\end{abstract}

\vspace{-0.3cm}
\section{Introduction}
With the upcoming bandwidth-hungry applications of the sixth generation (6G) networks such as virtual reality (VR) and holographic communications, ubiquitous and ultra-high-speed access are required \cite{wang2022ultra}. Utilizing the terahertz (THz) band will increase channel bandwidth and transmission capacity as it is considered one of the key enablers of 6G networks \cite{sarieddeen2020next}.
Compared to traditional radio frequency (RF) networks, dense THz networks have lower power consumption, smaller time delay, weaker radiation, better beam directivity, and higher interference immunity \cite{rappaport2013millimeter}. However, THz signals suffer from deep fading due to large free space path-loss and water molecule absorption, which significantly limit the effective communication distance \cite{chattopadhyay2015compact}. Hence, relay communications are used to cope with this limitation.

\par
Most of the literature focuses on coverage performance for a given relay link \cite{boulogeorgos2020outage,farrag2021outage,li2021performance}. {\color{black} The works in \cite{boulogeorgos2020outage,farrag2021outage} studied the outage probability, which is the complementary cumulative distribution function of the coverage probability, of dual-hop THz-THz links, and \cite{farrag2021outage} further optimized the transmission power.} In \cite{li2021performance}, the authors derived the outage probability and the ergodic capacity of dual-hop THz-RF links, respectively. However, the above distance-fixed three-point (source-relay-destination) model is simplified for a dense DF THz network.  {\color{black} The authors in \cite{sayehvand2020interference} analyzed the downlink coverage probability of the hybrid RF and THz network, but relay communication is not considered.}

Against this background, THz relay selection strategies are lacking in the literature. Moreover, since THz base stations are expected to coexist with RF base stations, a relay selection mechanism involving both THz and RF relays is necessary especially for ultra-dense networks (UDNs) and integrated access and backhaul (IAB) deployments. In this treatise, we propose a hybrid relay selection (HRS) protocol with RF and THz relays. 
Furthermore, the mathematical derivations are based on a realistic THz channel model that follows $\alpha-\mu$ which is an experimentally validated model \cite{papasotiriou2021experimentally}, and the relays are randomly located following a Poisson point process (PPP). The aim of this letter is to investigate the performance of the proposed protocol and how the relay selection is carried out when both RF and THz relays are involved in the selection process. Our proposed HRS protocol is compared with the optimal relay selection that requires a full and perfect instantaneous channel state information (CSI) of all nodes in the networks which is impractical to estimate. We also show when to use RF relays and/or THz relays. Finally, for the sake of completeness, we compare our protocol with RF and THz direct transmissions.

\vspace{-0.3cm}
\section{System Model}\label{section2}

\begin{figure}[t]
	\centering
	\includegraphics[width=0.75\linewidth]{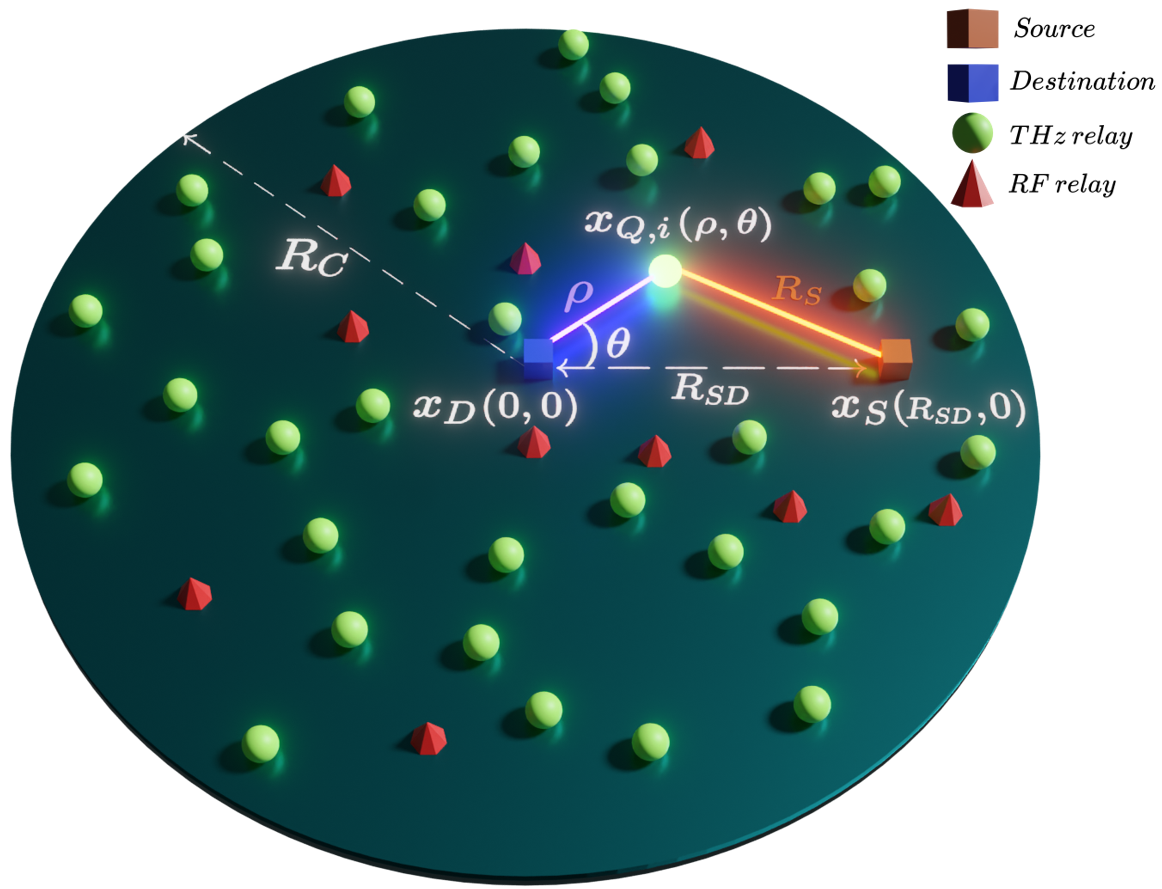}
	\caption{System model.}
	\label{fig:sys}
	\vspace{-6mm}
\end{figure}

\vspace{-0.1cm}
\subsection{Network Model}
We consider the system model depicted in Fig.~\ref{fig:sys}. Without loss of generality, we set the position of a destination ($D$) at the origin for tractability since the same performance is obtained for any other
$D$ locations due to Slivnyak's Theorem. Also, the direction from a source ($S$) to $D$ is the positive x-axis. In polar coordinate $\left(\rho, \theta \right)$, the positions of $S$ and $D$ can be denoted as $x_S(R_{SD},0)$ and $x_D(0,0)$, respectively, where $R_{SD}$ is the distance between $S$ and $D$. {\color{black} The locations of $Q$ relays form a homogeneous PPP denoted by $\Phi_Q=\{x_{Q,1},x_{Q,2},...\}$ and with density $\lambda_Q$, where $Q$ is replaced by RF in RF networks, and by THz in THz networks. } We assume $\Phi_{\rm{RF}}$ and $\Phi_{\rm{THz}}$ are independent. {\color{black}  All RF and THz relays, referred to as all nodes, are located on a two-dimensional circular disc $\mathcal{C}(x_D, R_C)$ centered at the origin with radius $R_C$. Note that the relays outside $\mathcal{C}(x_D, R_C)$ are not considered since they are far from $D$ and suffer a large signal attenuation.  }Hence, $\mathcal{C}(x_D, R_C)$ is equivalent to the entire ${\R}^{2}$. 
We assume that the locations of $S$, $D$, and the relays and the frequency band of the relays are shared, so the beams have been trained, and the communication targets are always within the main lobe of beams. {\color{black} Owing to the utilization of orthogonal sub-bands and narrow-beam transmission, it is extremely uncommon that the interference signal in the same sub-band is placed in the main lobe of the receiving device, hence the interference is minor compared to noise.}

\par
\par
\vspace{-0.3cm}
\subsection{Channel Model}
\subsubsection{RF Channel Model}
RF channels experience path-loss and small-scale fading; therefore, the corresponding received power can be expressed as, $h_{\rm{RF}}\left( R_X \right) = \varepsilon_{{\rm{RF}}}\,G_{{\rm{RF}}}\,\gamma_{\rm{RF}}\,R_X^{-\beta_{\rm{RF}}} \, \mathcal{X}_{\rm{RF}}$,
where $X = \{ S, D \}$, $R_X$ is the distance between a relay and $X$, $\varepsilon_{{\rm{RF}}}$ is the transmitting power, and $G_{{\rm{RF}}}$ is total antenna gain. 
The path-loss is modeled as $\gamma_{\rm{RF}} R_X^{-\beta_{\rm{RF}}}$, where $\beta_{\rm{RF}}$ is the path-loss exponent, $\gamma_{\rm{RF}}=\left({c}/{4 \pi \nu_{\rm{RF}} }\right)^2$, $c=3 \times 10^8$~m/s is the speed of light, and $\nu_{\rm{RF}}$ is the RF carrier frequency.
The small-scale fading $\mathcal{X}_{\rm{RF}}$ is subject to an exponential distribution with unit mean. The \ac{SNR} of the RF channel is given by
\vspace{-1mm}
\begin{sequation}
    {\rm{SNR}}_{{\rm{RF}},X} = \frac{\varepsilon_{{\rm{RF}}} \,G_{\rm{RF}} \gamma_{\rm{RF}} \, R_X^{-\beta_{\rm{RF}}} \, \mathcal{X}_{\rm{RF}}}{\sigma_{\rm{RF}}^2},
\vspace{-1mm}
\end{sequation}where $\sigma_{\rm{RF}}^2$ is the thermal noise. We note that $\sigma_{\rm{RF}}^2$ is a function of the transmission bandwidth $B_{\rm{RF}}$.

\par
\subsubsection{THz Channel Model}
According to \cite{sayehvand2020interference}, the received power in THz transmission is modeled as, $h_{\rm{THz}} \left( R_X \right) = \varepsilon_{{\rm{THz}}} \, G_{{\rm{THz}}}\, \gamma_{{\rm{THz}}} \, \mathcal{X}_{\rm{THz}}\, {\exp\left({-\beta_{\rm{THz}} R_X}\right)}/{R_X^2} $,
where $X = \{ S, D \}$, $R_X$ is the distance between a relay and $X$,  $\varepsilon_{{\rm{THz}}}$ is the transmitting power, and $G_{{\rm{THz}}}$ is the total antenna gain.
The path-loss is modeled as $\gamma_{{\rm{THz}}} {\exp\left({-\beta_{\rm{THz}} R_X}\right)}/{R_X^2}$, where $\gamma_{{\rm{THz}}}=\left({c}/{4 \pi \nu_{\rm{THz}} }\right)^2$ , $\nu_{\rm{THz}}$ is the THz carrier frequency in GHz,  and $\beta_{\rm{THz}}$ is the molecular absorption coefficient related to $\nu_{\rm{THz}}$. The small-scale fading $\mathcal{X}_{\rm{THz}}$ is modeled as $\alpha-\mu$ distribution which is an experimentally validated fading
model for THz frequencies \cite{papasotiriou2021experimentally}. 
Moreover, $\alpha$ denotes the fading parameter and $\mu$ denotes the normalized variance of the channel fading.
The \ac{CCDF} of $\mathcal{X}_{\rm{THz}}$ is given by $\bar{F}_{\mathcal{X}_{\rm{THz}}}(m) = {\Gamma(\mu,\mu m^{\frac{\alpha}{2}})}/{\Gamma(\mu)}$, where ${\Gamma(\mu,\mu m^{\frac{\alpha}{2}})}$ and ${\Gamma(\mu)}$ are the upper incomplete Gamma function and the Gamma function, respectively \cite{lei2017secrecy}. The \ac{SNR} of the THz channel is given by
\vspace{-1mm}
\begin{equation}
   {\rm{SNR}}_{{\rm{THz}},X}  = \frac{\varepsilon_{{\rm{THz}}} \,G_{{\rm{THz}}} \gamma_{{\rm{THz}}} \, \exp\left({-\beta_{\rm{THz}} R_X}\right) \, \mathcal{X}_{\rm{THz}}}{\sigma_{\rm{THz}}^2\, R_X^2},
\vspace{-1mm}
\end{equation}
where $\sigma_{\rm{THz}}^2$ is the thermal noise, which is a function of the
transmission bandwidth  $B_{\rm{THz}}$. 

\vspace{-0.3cm}
\subsection{Relay Selection Strategy}

%
Due to excessive path-loss and blockage, a half-duplex DF relay protocol is adopted since the direct link may be blocked.
In fixed topology DF networks, the optimal relay selection relies on the maximization of end-to-end ${\rm{SNR}}_{Q,X}$ or the maximization of the minimum ${\rm{SNR}}_{Q,X}$ of $S-x_{Q,i}$ and $x_{Q,i}-D$ links. However, this protocol requires a full and perfect instantaneous CSI of all the nodes in the networks, that is, CSI between $S$ and all relays, and between all relays and $D$. In addition, due to the randomness of the relay locations, analysis is intractable \cite{wang2022stochastic}. 
      {\color {black} 
      Therefore, we proposed a slightly suboptimal protocol that only requires instantaneous CSI between $S$ and all nodes while still being mathematically tractable. } 
The protocol has the following selection strategy:
we first select two sets of relays $\widetilde{\Phi}_Q$ that can meet the attempted rate $y_{th}$ of $S$. Among these sets, only the relay $x_{\rm{RF},i} \in \widetilde{\Phi}_{\rm{RF}}$ or $x_{\rm{THz},i} \in \widetilde{\Phi}_{\rm{THz}}$ that has the
best link to $D$, that is, provides the maximum average rate with $D$ is selected \cite{sayehvand2020interference}. The set of relays that can establish reliable communication with $S$ can be expressed as
\vspace{-1mm}
\begin{equation}
    \widetilde{\Phi}_Q=\{x_{Q,i} \in \Phi_Q, \, {\rm{SNR}}_{Q,S} \geq \tau_Q \}, \ Q = \{ \rm{RF}, \rm{THz} \},
\vspace{-1mm}
\end{equation}
where $\tau_Q$ is a predefined threshold, ${\rm{SNR}}_{Q,S}$ is the SNR of S-relay link.  Subsequently, the relation between achievable rate $y_{Q,X}$ and ${\rm{SNR}}_{Q,X}$  is given by
\vspace{-1mm}
\begin{equation}
  y_{Q,X} = \frac{B_{Q}}{2} \log_{2}\left(1+{\rm{SNR}}_{Q,X}\right), \vspace{-1mm}  
\end{equation}
where $B_{Q}$ is the transmission bandwidth of $Q$ link and the factor 1/2 is used because two time slots are required to
transmit data via the relay node. 
    {\color{black} The rate coverage probability is defined as the probability of achieving the desired rate $y_{th}$, i.e.,   ${\textrm{SNR}}_{Q,X}>\tau_Q = 2^{\left(2y_{th}/B_Q\right)}-1$.}




\vspace{-0.2cm}
\section{{\color{black}Coverage Probability of HRS Protocol}}
In this section, we derive the coverage probability, the intermediate distance distributions, and the association probabilities of the HRS protocol.  Note that the nearest relay provides the largest average rate for $D$ in our model, so the distance distributions are given in Lemmas. 
\vspace{-1mm}
{\color{black}
\begin{lemma}\label{lemma1}
The complementary cumulative distribution function (\ac{CCDF}) of the distance between D and its nearest RF relay is given by
\vspace{-1mm}
\begin{sequation}\label{eq.ccdf.rf}
\begin{split}
    \overline{F}_{R_D,{\rm{RF}}} (r) = \exp \bigg( - \int_0^r \int_0^{2\pi} \lambda_{\rm{RF}} \exp \Big( -\tau_{\rm{RF}} \varepsilon_{{\rm{RF}}}^{-1} \\
    \times G_{{\rm{RF}}}^{-1} \gamma_{{\rm{RF}}}^{-1} \!\left( \rho^2 \!+ \!R_{SD}^2 \!-\! 2\rho R_{SD} \cos\theta \right)^{\frac{\beta_{\rm{RF}}}{2}} \sigma_{\rm{RF}}^2 \Big) \bigg) \rho \, \mathrm{d}\theta \mathrm{d} \rho.
\end{split}
\vspace{-1mm}
\end{sequation}
where $0 < r \leq R_C$.

Then, the \ac{PDF} of the distance between D and its nearest RF relay is given by,
\vspace{-1mm}
\begin{sequation}\label{PDF of RF}
\begin{split}
    & f_{R_D,{\rm{RF}}}  (r) = r \lambda_{\rm{RF}} \int_0^{2\pi} \exp \Big( -\tau_{\rm{RF}} \varepsilon_{{\rm{RF}}}^{-1}G_{{\rm{RF}}}^{-1}  \gamma_{{\rm{RF}}}^{-1} \\
    &\times \left( \rho^2 + R_{SD}^2 - 2\rho R_{SD} \cos\theta \right)^{\frac{\beta_{\rm{RF}}}{2}} \sigma_{\rm{RF}}^2 \Big) \mathrm{d}\theta \\
    &\times \exp \bigg( - \int_0^r \int_0^{2\pi} \lambda_{\rm{RF}} \exp \Big( -\tau_{\rm{RF}} \varepsilon_{{\rm{RF}}}^{-1}G_{{\rm{RF}}}^{-1}  \gamma_{{\rm{RF}}}^{-1} \\
    &\times \left( \rho^2 + R_{SD}^2 - 2\rho R_{SD} \cos\theta \right)^{\frac{\beta_{\rm{RF}}}{2}} \sigma_{\rm{RF}}^2 \Big) \bigg) \rho \, \mathrm{d}\theta \mathrm{d} \rho.\\
\end{split}
\vspace{-1mm}
\end{sequation}
\begin{proof}
The CCDF of the distance can be derived by using the void probability of a homogeneous PPP \cite{haenggi2012stochastic}
\begin{sequation}\label{app1_1}
\begin{split}
    \overline{F}_{R_D,{\rm{RF}}} (r)  =  \mathbb{P} \left[{\rm No\,relays\,in\,} \mathcal{C}(x_D,r)  \right] = \exp\left( - \Lambda_{\rm{RF}}\left( r \right) \right),
    \end{split}
\end{sequation}where $\mathcal{C}(x_D,r)$ is the disk centered at the origin ($x_D$) with radius $r$.
The mean number of RF relays $x_{\rm{RF},i} \in \widetilde{\Phi}_{\rm{RF}}$ in the circular disc $\mathcal{C}(x_D,r)$, denoted by $\Lambda_{\rm{RF}}\left( r \right)$, can be written as
\vspace{-1mm}
\begin{equation}\label{app1_2}
    \Lambda_{\rm{RF}}\left( r \right) = \int_0^r \int_0^{2\pi} \widehat{\lambda}_{\rm{RF}}(\rho,\theta) \rho \, \mathrm{d}\theta \mathrm{d} \rho,
\vspace{-1mm}
\end{equation}where $\rho$ and $\theta$ are the radial distance and polar angle in the polar coordinate system respectively, and $\widehat{\lambda}_{\rm{RF}}(\rho,\theta)$ is the density of $\widetilde{\Phi}_{\rm{RF}}$. The set $\widetilde{\Phi}_{\rm{RF}}$ can be regarded as a dependent-thinning of $\Phi_{\rm{RF}}$, and only the relays that have an ${\rm{SNR}}_{{\rm{RF}},S}$ greater than the threshold $\tau_{\rm{RF}}$ are retained. Hence, the set $\widetilde{\Phi}_{\rm{RF}}$ is a inhomogeneous PPP of
density $\widehat{\lambda}_{\rm{RF}}(\rho,\theta)$. The density $\widehat{\lambda}_{\rm{RF}}(\rho,\theta)$ can be derived as follows
\vspace{-1mm}
\begin{sequation}\label{app1_3}
\begin{split}
\widehat{\lambda}_{\rm{RF}}(\rho,\theta) &= \lambda_{\rm{RF}} \, \mathbb{P} \left[\frac{\varepsilon_{{\rm{RF}}} \, G_{\rm{RF}} \gamma_{\rm{RF}} \, R_{S}^{-\beta_{\rm{RF}}} \, \mathcal{X}_{\rm{RF}}}{\sigma_{\rm{RF}}^2} > \tau_{\rm{RF}}\right] \\
    &= \lambda_{\rm{RF}} \, \mathbb{P} \left[ \mathcal{X}_{\rm{RF}} > \tau_{\rm{RF}} \varepsilon_{{\rm{RF}}}^{-1} G_{{\rm{RF}}}^{-1} \gamma_{{\rm{RF}}}^{-1} R_{S} ^{\beta_{\rm{RF}}} \sigma_{\rm{RF}}^2 \right] \\
    &= \lambda_{\rm{RF}} \exp\left( -\tau_{\rm{RF}} \varepsilon_{{\rm{RF}}}^{-1} G_{{\rm{RF}}}^{-1} \gamma_{{\rm{RF}}}^{-1} R_{S} ^{\beta_{\rm{RF}}} \sigma_{\rm{RF}}^2 \right),
\end{split}
\vspace{-2mm}
\end{sequation}where $R_S$ is the distance between $S$ and relay and can expressed as
\vspace{-2mm}
\begin{equation}\label{app1_4}
    R_{S}^2 = \rho^2 + R_{SD}^2 - 2\rho \, R_{SD} \cos\theta.
\vspace{-1mm}
\end{equation}

By substituting (\ref{app1_2}), (\ref{app1_3}), and (\ref{app1_4}) into (\ref{app1_1}), we obtain (\ref{eq.ccdf.rf}). Then, using $f_{R_D,{\rm{RF}}}(r) = -\frac{\mathrm{d}}{\mathrm{d}r}\overline{F}_{R_D,{\rm{RF}}} (r)$ and Leibniz integral rule, (\ref{PDF of RF}) is derived straightforwardly.

\end{proof}
\end{lemma}
\vspace{-1mm}
\begin{lemma}\label{lemma3}
The \ac{CCDF} and PDF of the distance between D and its nearest THz relay are given by (\ref{CCDF of THz}) and (\ref{PDF of THz}), respectively, shown at the top of the next page.

\begin{table*}
\vspace{-6mm}
\begin{sequation}\label{CCDF of THz}
\begin{split}
    \overline{F}_{R_D,{\rm{THz}}} (r) = \exp \left( - \frac{\lambda_{\rm{THz}}}{\Gamma\left(\mu\right)} \int_0^r \int_0^{2\pi} {\Gamma \left(\mu, \mu \, \left(  \frac{\tau_{\rm{THz}} \left(\rho^2 + R_{SD}^2 - 2\rho R_{SD} \cos\theta\right) \sigma_{\rm{THz}}^2 } {\varepsilon_{{\rm{THz}}}G_{{\rm{THz}}} \gamma_{{\rm{THz}}} \exp\left(-\beta_{\rm{THz}} \left( \rho^2 + R_{SD}^2 - 2\rho R_{SD} \cos\theta \right)^{\frac{1}{2}} \right)} \right)^{\frac{\alpha}{2}} \right)} \rho \, \mathrm{d}\theta \mathrm{d} \rho \right).
\end{split}
\end{sequation}
\vspace{-0.3cm}
\begin{sequation}\label{PDF of THz}
\begin{split}
   f_{R_D,{\rm{THz}}} (r) &= r \, \frac{\lambda_{\rm{THz}}}{\Gamma\left(\mu\right)}  \int_0^{2\pi} {\Gamma \left(\mu, \mu \, \left(  \frac{\tau_{\rm{THz}} \left(r^2 + R_{SD}^2 - 2r R_{SD} \cos\theta\right) \sigma_{\rm{THz}}^2 } {\varepsilon_{{\rm{THz}}}G_{{\rm{THz}}} \gamma_{{\rm{THz}}} \exp\left(-\beta_{\rm{THz}} \left( r^2 + R_{SD}^2 - 2r R_{SD} \cos\theta \right)^{\frac{1}{2}} \right)} \right)^{\frac{\alpha}{2}} \right)} \, \mathrm{d}\theta \\ 
   \vspace{-1mm}
   &\times \exp \left( - \frac{\lambda_{\rm{THz}}}{\Gamma\left(\mu\right)} \int_0^r \int_0^{2\pi} {\Gamma \left(\mu, \mu \, \left(  \frac{\tau_{\rm{THz}} \left(\rho^2 + R_{SD}^2 - 2\rho R_{SD} \cos\theta\right) \sigma_{\rm{THz}}^2 } {\varepsilon_{{\rm{THz}}}G_{{\rm{THz}}} \gamma_{{\rm{THz}}} \exp\left(-\beta_{\rm{THz}} \left( \rho^2 + R_{SD}^2 - 2\rho R_{SD} \cos\theta \right)^{\frac{1}{2}} \right)} \right)^{\frac{\alpha}{2}} \right)} \rho \, \mathrm{d}\theta \mathrm{d} \rho \right).
\end{split}
\vspace{-0.2cm}
\end{sequation}
\vspace{-0.1cm}
\noindent\rule{\linewidth}{0.2mm}
\vspace{-0.4cm}
\end{table*}

\begin{proof}
Following the same steps as in Lemma~\ref{lemma1}, we obtain
\vspace{-1mm}
\begin{sequation}\label{app2_1}
    \overline{F}_{R_D,{\rm{THz}}} (r) = \exp \left( {-\int_0^r \int_0^{2\pi} \widehat{\lambda}_{\rm{THz}}(\rho,\theta) \rho \, \mathrm{d}\theta \mathrm{d} \rho} \right),
\vspace{-1mm}
\end{sequation}where $\widehat{\lambda}_{\rm{THz}}(\rho,\theta)$ is the density of $\widetilde{\Phi}_{\rm{THz}}$, which is given by
\vspace{-1mm}
\begin{sequation}\label{app2_2}
\begin{split}
    &\widehat{\lambda}_{\rm{THz}}(\rho,\theta)  
    = \lambda_{\rm{THz}} \, \mathbb{P} \left[ \frac{\varepsilon_{{\rm{THz}}} \,G_{{\rm{THz}}} \gamma_{{\rm{THz}}} \, \, \mathcal{X}_{\rm{THz}}}{\sigma_{\rm{THz}}^2\,\exp\left({\beta_{\rm{THz}} R_S}\right)\, R_S^2}> \tau_{\rm{THz}}\right] \\
    &= \lambda_{\rm{THz}} \, \mathbb{P} \bigg[ \mathcal{X}_{\rm{THz}} > \frac{\tau_{\rm{THz}} R_{S}^2 \exp \left( \beta_{\rm{THz}} R_{S} \right) \sigma_{\rm{THz}}^2}{\varepsilon_{{\rm{THz}}}G_{{\rm{THz}}} \gamma_{{\rm{THz}}}}  \bigg] \\
    &\overset{(a)}{=} \frac{\lambda_{\rm{THz}}}{\Gamma\left(\mu\right)} \Gamma \left(\mu, \mu \, \left(  \frac{\tau_{\rm{THz}} R_{S}^2 \exp\left(\beta_{\rm{THz}} R_{S} \right) \sigma_{\rm{THz}}^2 }  {\varepsilon_{{\rm{THz}}}G_{{\rm{THz}}} \gamma_{{\rm{THz}}}} \right)^{\frac{\alpha}{2}} \right),
\end{split}
\vspace{-1mm}
\end{sequation}where (a) follows the definition of the \ac{CCDF} of $\mathcal{X}_{\rm{THz}}$. Substituting (\ref{app2_2}) and (\ref{app1_4}) into (\ref{app2_1}) concludes the proof of (\ref{CCDF of THz}). Similar to the proof of (\ref{PDF of RF}) in Lemma~\ref{lemma1}, (\ref{CCDF of THz}) is given by $f_{R_D,{\rm{THz}}}(r) = -\frac{\mathrm{d}}{\mathrm{d}r}\overline{F}_{R_D,{\rm{THz}}} (r)$.
\end{proof}
\end{lemma}}

Using the above Lemmas, we derive the distribution of RF and THz relays that are closest to $D$. However, for the nearest RF relay to be selected, it is necessary to ensure that the achievable rate of THz transmission is less than that of RF transmission, and vice versa. Therefore, we calculate the (\textbf{selection}) association probabilities in the following two Lemmas.


\vspace{-1mm}
\begin{lemma}\label{lemma5}
Given that the distance between $D$ and its nearest RF relay is $r$, the probability that $D$ \textbf{selects}/associates with this RF relay is given by
\vspace{-1mm}
\begin{equation}\label{association of RF}
    P_{\rm{RF}}^A (r) = 1 - \int_0^{R_{\rm{R2T}}(r)} f_{R_D,{\rm{THz}}}(\rho) \mathrm{d} \rho,
\vspace{-1mm}
\end{equation}
where $R_{\rm{R2T}}$ is given by
\vspace{-1mm}
\begin{sequation}\label{R2T}
\begin{split}
   & R_{\rm{R2T}}(r) = \frac{2}{\beta_{\rm{THz}}} \times \\
   & \mathcal{W}\!\!\left(\!\!\frac{\beta_{\rm{THz}}}{2}\! \!\left(\!\!\!\left(\!\!\left(\! 1\!\!+\! \frac{\varepsilon_{{\rm{RF}}} G_{{\rm{RF}}}\gamma_{{\rm{RF}}} }{r^{\beta_{\rm{RF}}}\sigma_{{\rm{RF}}}^2}\! \right)^{\!\!\frac{B_{{\rm{RF}}}}{B_{{\rm{THz}}}}}\!\!\!\!\!-\!1\!\!\right)\!\! \frac{\sigma_{{\rm{THz}}}^2}{\varepsilon_{{\rm{THz}}} G_{{\rm{THz}}}\gamma_{{\rm{THz}}}}\!\!\right)^{\!\!\!-\!\frac{1}{2}}\!\right)\!.
\end{split}
\vspace{-1mm}
\end{sequation}
We note that $\mathcal{W}\left( \cdot \right)$ is the Lambert $\mathcal{W}$-function defined as the inverse of the function $f_{\mathcal{W}}(m)=m e^m$. 
\begin{proof}

The distance of the nearest THz relay, denoted by $R_{\rm{R2T}}$, can be solved by setting the average achievable rate of the RF relay with distance $r$ to $D$ equals to that of the THz relay with distance $R_{\rm{R2T}}$ to $D$ \cite{UAV_SG}
\vspace{-1mm}
\begin{equation}
\begin{split}
    \mathbb{E}_{\mathcal{X}_{\rm{RF}}} & \left[{\rm{SNR}}_{{\rm{RF}},D}|R_{{\rm{RF}},D}=r\right] \\
    &=\mathbb{E}_{\mathcal{X}_{\rm{THz}}}  \left[{\rm{SNR}}_{{\rm{THz}},D}|R_{{\rm{THz}},D}=R_{\rm{R2T}}\right].
\end{split}
\vspace{-1mm}
\end{equation}

From the above equation, the relationship between $R_{\rm{R2T}}$ and $r$ is given by (\ref{R2T}). By the definition of association probability $P_{\rm{RF}}^A (r)$ in Lemma~\ref{lemma5} and the maximum average received power associated strategy, the following result is given
\vspace{-1mm}
\begin{equation}
\begin{split}
    P_{\rm{RF}}^A (r) &=\mathbb{P} \left[{\rm No\,relays\,in\,} \mathcal{C}(x_D,R_{\rm{R2T}})  \right] =
    \\
    &= 1 - \int_0^{R_{\rm{R2T}}(r)} f_{R_D,{\rm{THz}}}(\rho) \mathrm{d} \rho.
\end{split}
\vspace{-1mm}
\end{equation}
This concludes the proof of Lemma~\ref{lemma5}.
\end{proof}
\end{lemma}

\vspace{-1mm}
\begin{lemma}\label{lemma6}
Given that the distance between $D$ and its nearest THz relay is $r$, the probability that $D$ \textbf{selects}/associates with this THz relay is given by
\vspace{-1mm}
\begin{equation}\label{association of THz}
    P_{\rm{THz}}^A (r) = 1 - \int_0^{R_{\rm{T2R}}(r)} f_{R_D,{\rm{RF}}}(\rho) \mathrm{d} \rho,
\vspace{-1mm}
\end{equation}
where $f_{R_D,{\rm{RF}}}(r)$ is defined in (\ref{PDF of RF}) and $R_{\rm{T2R}}(r)$ is given by
\vspace{-1mm}
\begin{sequation}
\begin{split}
   & R_{\rm{T2R}}(r) = \\
   &\left(\!\!\!\left(\!\!\!\left(\!\!1\!\!+\!\! \frac{\varepsilon_{{\rm{THz}}}G_{\rm{THz}} \gamma_{\rm{THz}}}{r^2 \exp\left(\beta_{\rm{THz}}r \right) \sigma_{\rm{THz}}^2}\!\right)^\frac{{B_{\rm{THz}}}}{{B_{\rm{RF}}}}\!\!\!\!\!-\!1\!\!\right)\!\frac{\sigma_{\rm{RF}}^2}{\varepsilon_{{\rm{RF}}}G_{\rm{RF}} \gamma_{\rm{RF}}} \!\!\right)^{-\frac{1}{\beta_{\rm{RF}}}}\!.
\end{split}
\vspace{-1mm}
\end{sequation}
\begin{proof}
The proof of Lemma~\ref{lemma6} is similar to that of Lemma~\ref{lemma5} and therefore, omitted here.
\end{proof}
\end{lemma}

\begin{figure*}[htbp]
\vspace{-0.2cm}
\begin{minipage}[t]{0.32\linewidth}
\centering
\includegraphics[width=0.82\linewidth]{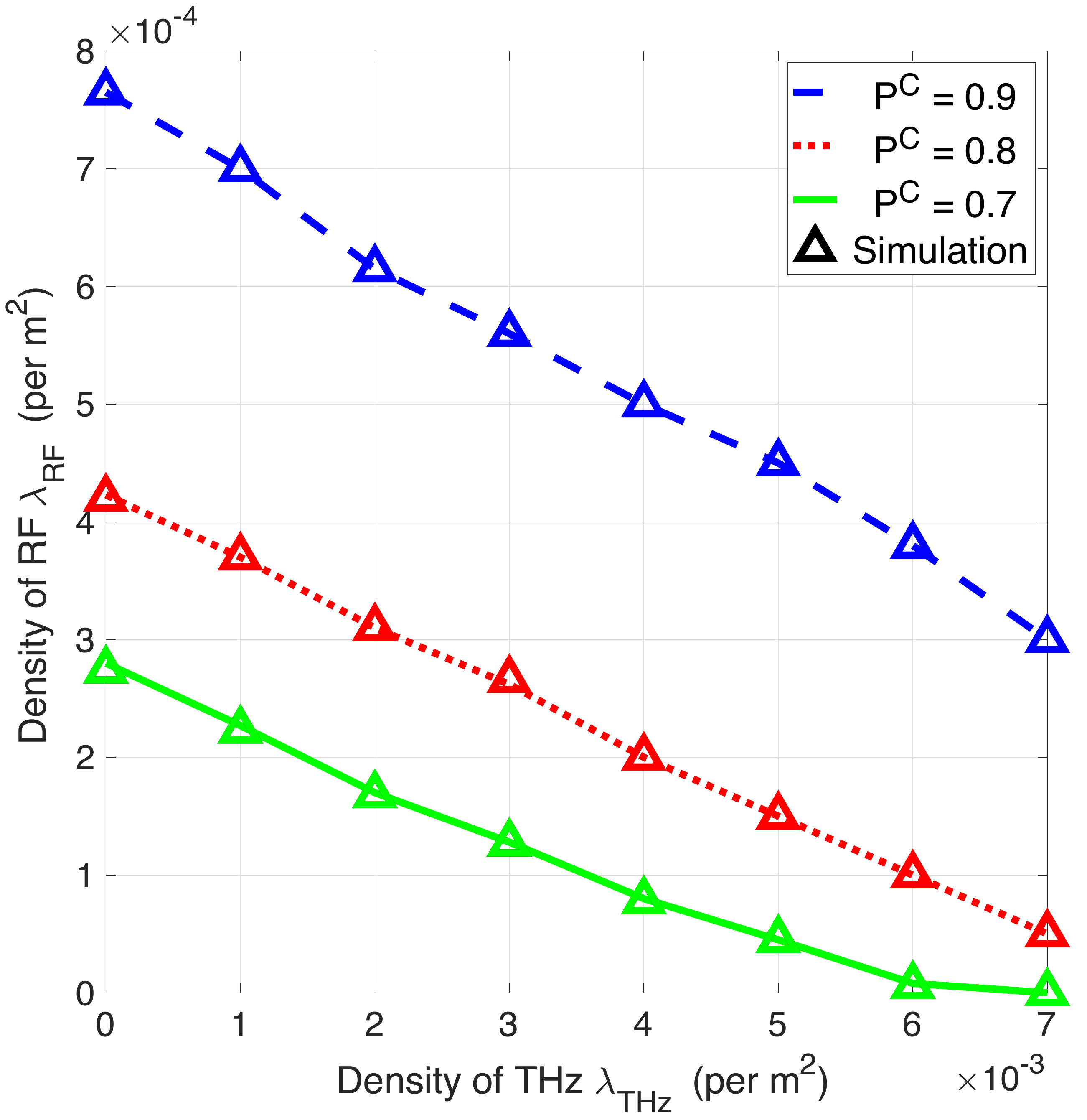}
\vspace{-0.1cm}
\caption{Density of RF and THz relays under the same coverage probability of HRS.}
\label{fig1}
\end{minipage}
\hfill
\begin{minipage}[t]{0.32\linewidth}
\centering
\includegraphics[width=0.82\linewidth]{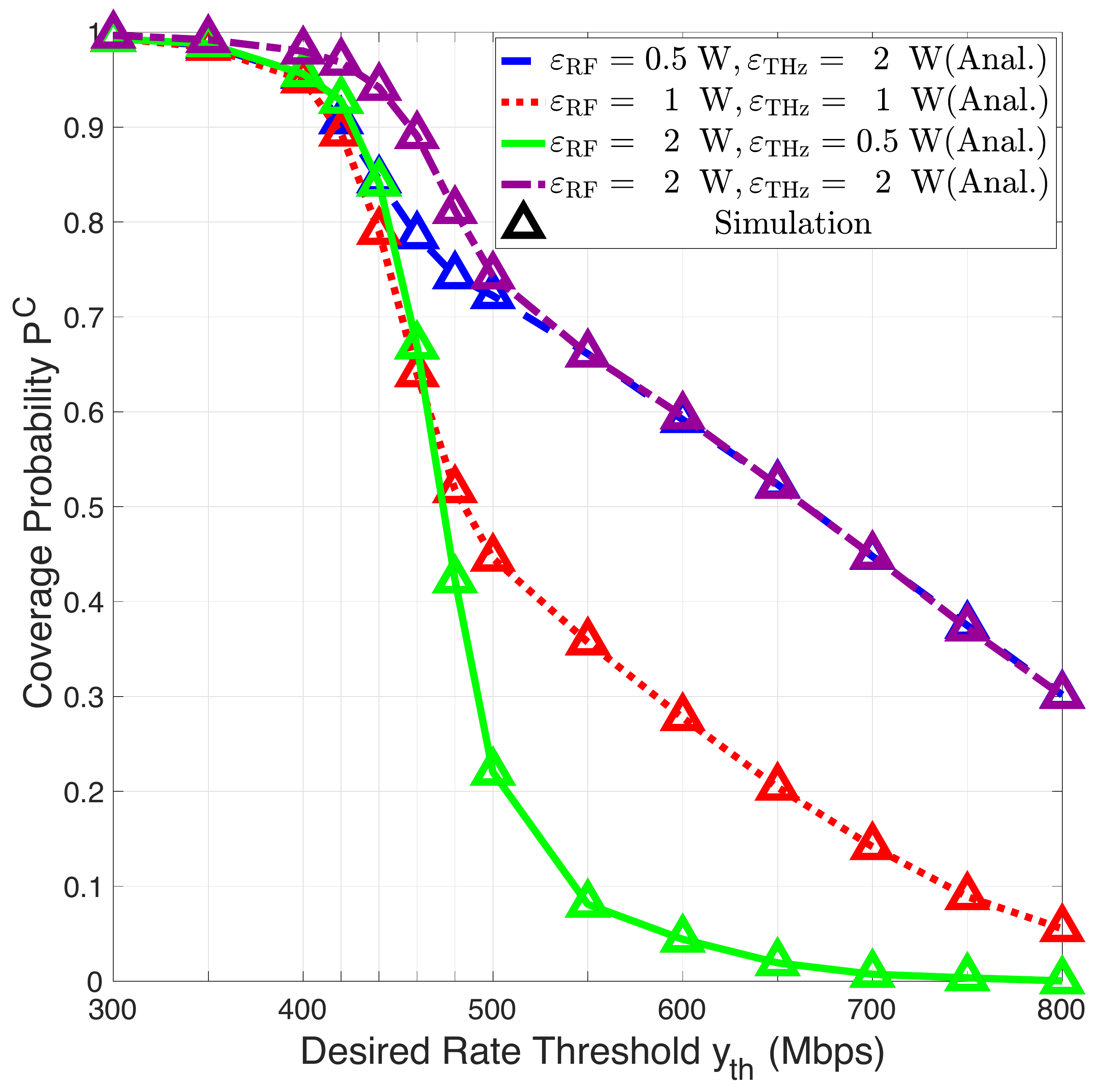}
\vspace{-0.1cm}
\caption{Impact of transmission power and desired rate threshold on HRS.}
\label{fig2}
\end{minipage}
\hfill
\begin{minipage}[t]{0.32\linewidth}
\centering
\includegraphics[width=0.82\linewidth]{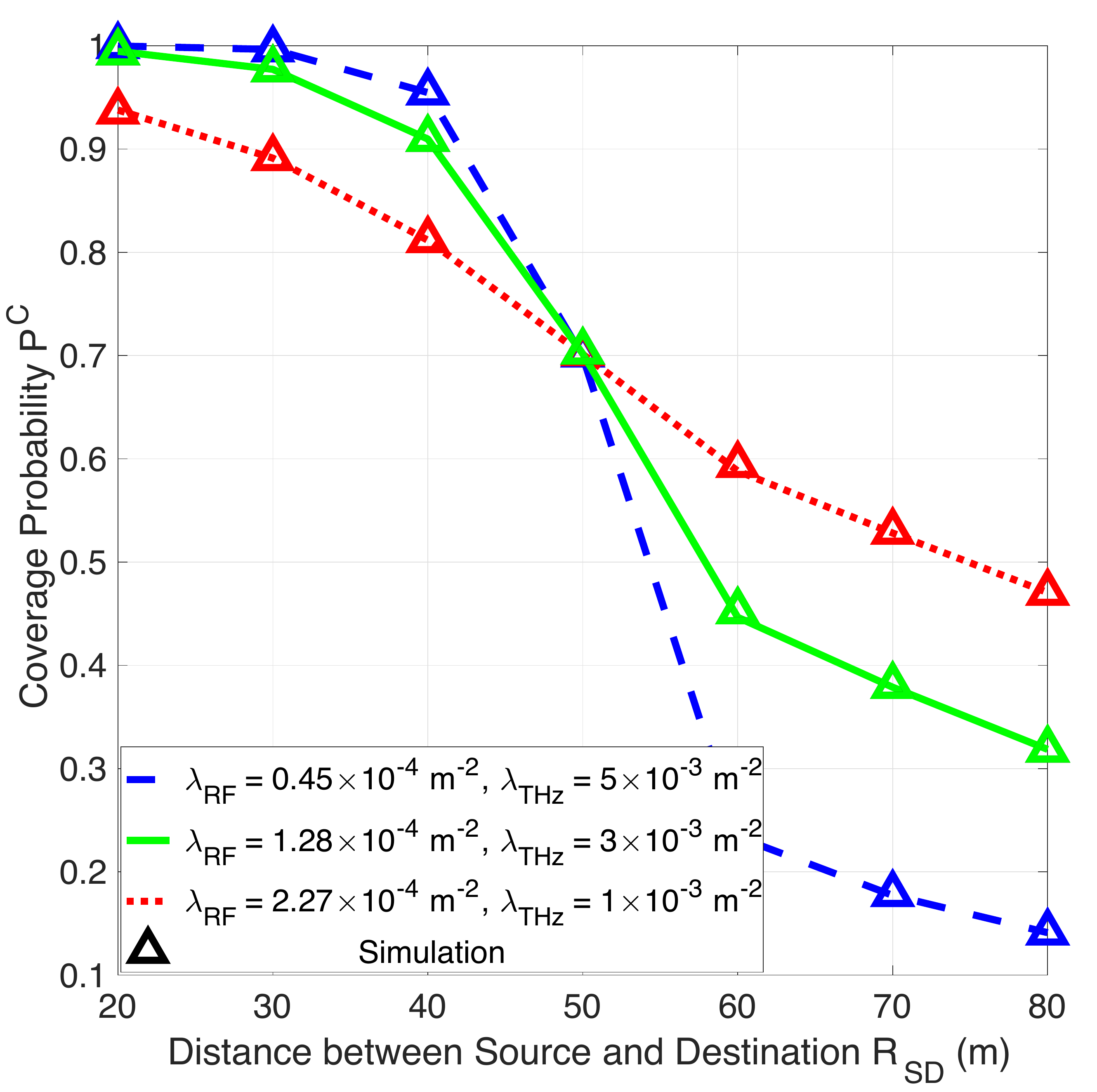}
\vspace{-0.1cm}
\caption{Impact of relay density and source-destination distance on HRS.}
\label{fig3}
\end{minipage}
\vspace{-0.6cm}
\end{figure*}

Based on the distance distributions and association probabilities, the main result of this letter is given by the following theorem.
\vspace{-1mm}
\begin{theorem}\label{theorem}
The coverage probability at $D$ using the HRS protocol is given by
\vspace{-1mm}
\begin{sequation}\label{coverage probability}
\begin{split}
    P^C =& \int_0^{R_C} f_{R_D,{\rm{RF}}}(\rho) \exp\left( - \frac{\tau_{\rm{RF}} \rho^{\beta_{\rm{RF}}} \sigma_{\rm{RF}}^2}{\varepsilon_{{\rm{RF}}}G_{{\rm{RF}}}\gamma_{{\rm{RF}}}} \right) P_{\rm{RF}}^A\left(\rho\right) \, \mathrm{d}\rho \\
    &+ \int_0^{R_C} f_{R_D,{\rm{THz}}}(\rho) P_{\rm{THz}}^A\left(\rho\right) \frac{1}{\Gamma \left(\mu\right)} \\
    &\times {\Gamma \! \left(\!\mu, \mu \, \left(  \frac{\tau_{\rm{THz}} \rho^2 \sigma_{\rm{THz}}^2 } {\varepsilon_{{\rm{THz}}}G_{{\rm{THz}}}  \gamma_{{\rm{THz}}}  \exp\left(-\beta_{\rm{THz}} \rho \right)} \right)^{\frac{\alpha}{2}} \right)}  \, \mathrm{d} \rho.
\end{split}
\vspace{-1mm}
\end{sequation}

\begin{proof}
{\color{black}  The coverage probability $P^C$ is defined as the probability that the received ${\rm{SNR}}_{Q,D}$ at the receiver $D$ from the selected $Q$ relay is above the predefined threshold $\tau_Q$, i.e., $P^C = \mathbb{P}\left[{\rm{SNR}}_{Q,D}>\tau_Q\right]$.} We divide the overall coverage probability $P^C$ into coverage probability associated with RF relay $P_{\rm{RF}}^C$ and coverage probability associated with THz relay $P_{\rm{THz}}^C$, i.e., $P^C=P_{\rm{RF}}^C+P_{\rm{THz}}^C$. Furthermore, $P_{\rm{RF}}^C$ can be obtained by (\romannumeral1) solving the coverage probability given that the distance of $S-x_{\rm{RF},i}$ is $R_{{\rm{RF}},D}=\rho$; (\romannumeral2) multiplying the coverage probability with the association probability given that the distance of $S-x_{\rm{RF},i}$ is $R_{{\rm{RF}},D}=\rho$ (which is denoted as $P_{\rm{RF}}^A\left(\rho\right)$); (\romannumeral3) compute the expectation $\mathbb{E}_{\rho}$. Hence, $P_{\rm{RF}}^C$ is calculated as follows
\begin{sequation}
\begin{split}
    P_{\rm{RF}}^C &= \mathbb{E}_{\rho}\left[  \mathbb{P}\left[ {\rm{SNR}}_{{\rm{RF}},D}>\tau_{\rm{RF}} | R_{{\rm{RF}},D}=\rho \right] P_{\rm{RF}}^A\left(\rho\right) \right] \\
    &=  \mathbb{E}_{\rho}\left[ \exp\left( - \frac{\tau_{\rm{RF}} \rho^{\beta_{\rm{RF}}} \sigma_{\rm{RF}}^2}{\varepsilon_{{\rm{RF}}}G_{{\rm{RF}}}\gamma_{{\rm{RF}}}} \right) P_{\rm{RF}}^A\left(\rho\right) \right] \\
    &= \int_0^{R_C} \!\!\!f_{R_D,{\rm{RF}}}(\rho) \exp\left( - \frac{\tau_{\rm{RF}} \rho^{\beta_{\rm{RF}}} \sigma_{\rm{RF}}^2}{\varepsilon_{{\rm{RF}}}G_{{\rm{RF}}}\gamma_{{\rm{RF}}}} \right) P_{\rm{RF}}^A\left(\rho\right) \, \mathrm{d}\rho. 
\end{split}
\end{sequation}The proof of $P_{\rm{THz}}^C$ is similar to the proof of $P_{\rm{RF}}^C$, therefore, it is omitted here.
{\color{black} We note that the association probability $P_{Q}^A$ is contained in the coverage probability associated with $Q$ relay $P_{Q}^C$.}
By using $P^C=P_{\rm{RF}}^C+P_{\rm{THz}}^C$, the overall coverage probability is derived.
\end{proof}
\end{theorem}

\vspace{-5mm}
\section{Numerical Results} \label{Num}
In this section, we provide selected simulation results to demonstrate the performance of the proposed HRS protocol. The simulations coincide perfectly with the theoretical analysis, and 
 each simulation is performed over $10^6$ independent network realizations within a circular disc of radius $R_C = 200$~m.
We consider the THz link with an antenna gain $G_{\rm{THz}} =40$~dBi and a carrier frequency $\nu_{\rm{THz}} = 1.8$~THz. The corresponding absorption value $\beta_{\rm{THz}}$ is chosen from the realistic database, which is $0.2$~m$^{-1}$ \cite{GORDON20173}. To simulate the $\alpha - \mu$ channel, we set $\alpha = 2$ and $\mu = 4$ for the THz link. The carrier frequency $\nu_{\rm{RF}}$ of the RF link is $2.1$~GHz. The antenna gain $G_{\rm{RF}}$ is $20$~dBi and path-loss exponent $\beta_{\rm{RF}} = 2.5$. {\color{black} We set the noise power to $-174$~dBm/Hz for both links such as $\sigma^{2}_{Q}=-174+10\log_{10} B_Q$ \rm{dBm} 
, while the transmission bandwidths $B_{\rm{RF}} = 40$~MHz and $B_{\rm{THz}} = 0.5$~GHz, by calculation $\sigma^{2}_{\rm{RF}} = -98$~dBm and $\sigma^{2}_{\rm{THz}} = -87$~dBm. }


{\color{black} Fig.~\ref{fig1} shows the density of RF and THz relays in the HRS protocol required to achieve a given coverage probability when $\varepsilon_{{Q}}=$ $1$~W, $R_{SD} = 50$~m, and $y_{th}= 420$~Mbps. For a given coverage probability, when $\lambda_{\rm{THz}}$ increases linearly, $\lambda_{\rm{RF}}$ decreases roughly linearly. Moreover, when the coverage probability grows from $0.7$ to $0.9$, the slopes of the three curves increase consecutively, but the amplitude of the increase is quite small. Specifically,  approximately $20$ additional THz relays are required for $1$ RF relay reduced to maintain the same coverage probability for an area of $2 \times 10^4$ m$^2$. }

\begin{figure*}
    \centering
    \vspace{-0.9cm}
    \subfigure[$R_{SD} = 20$~m] {\includegraphics[width=0.27\textwidth]{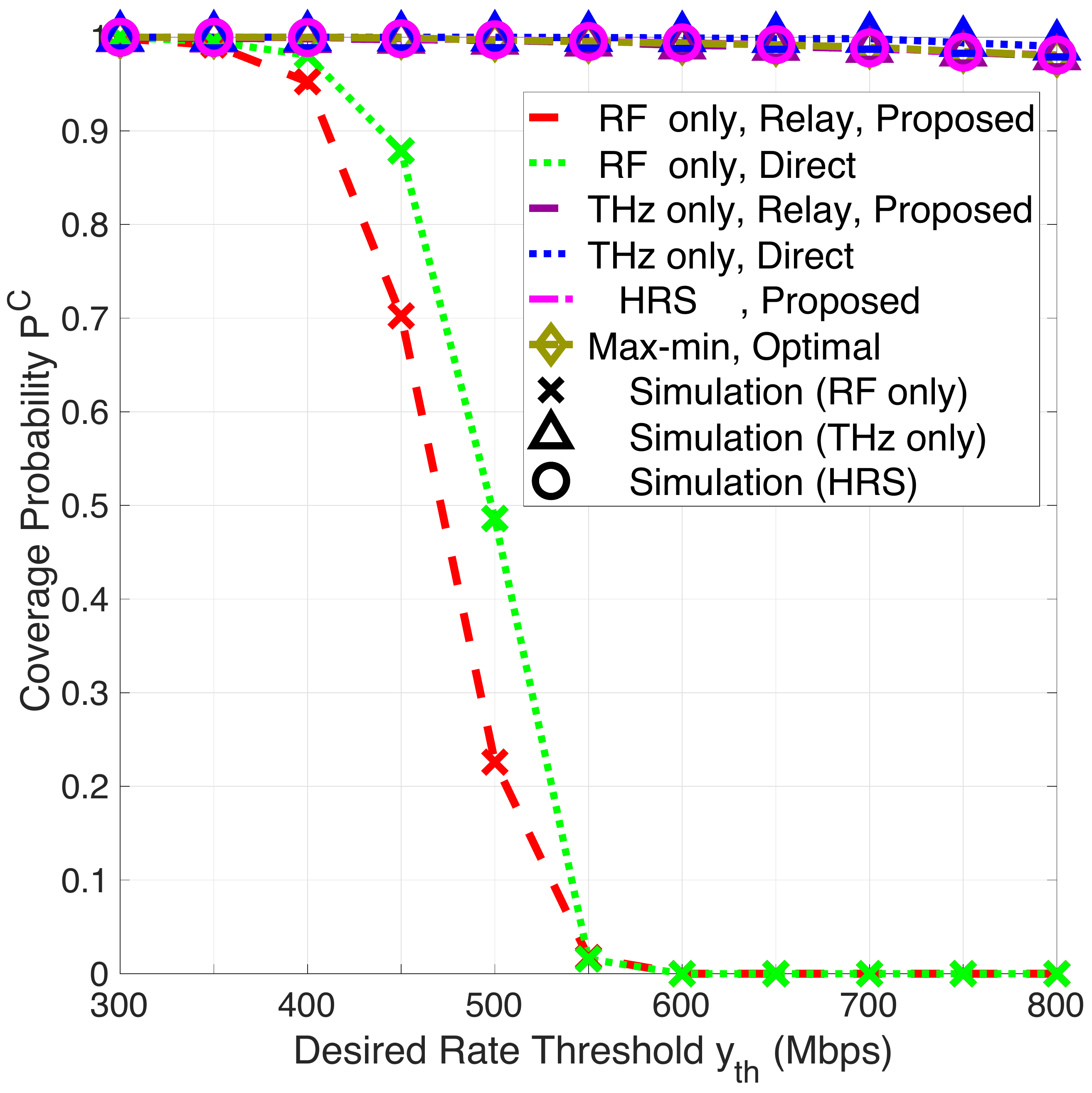}} 
    \hspace{0.4cm}
    \subfigure[$R_{SD} = 50$~m]{\includegraphics[width=0.27\textwidth]{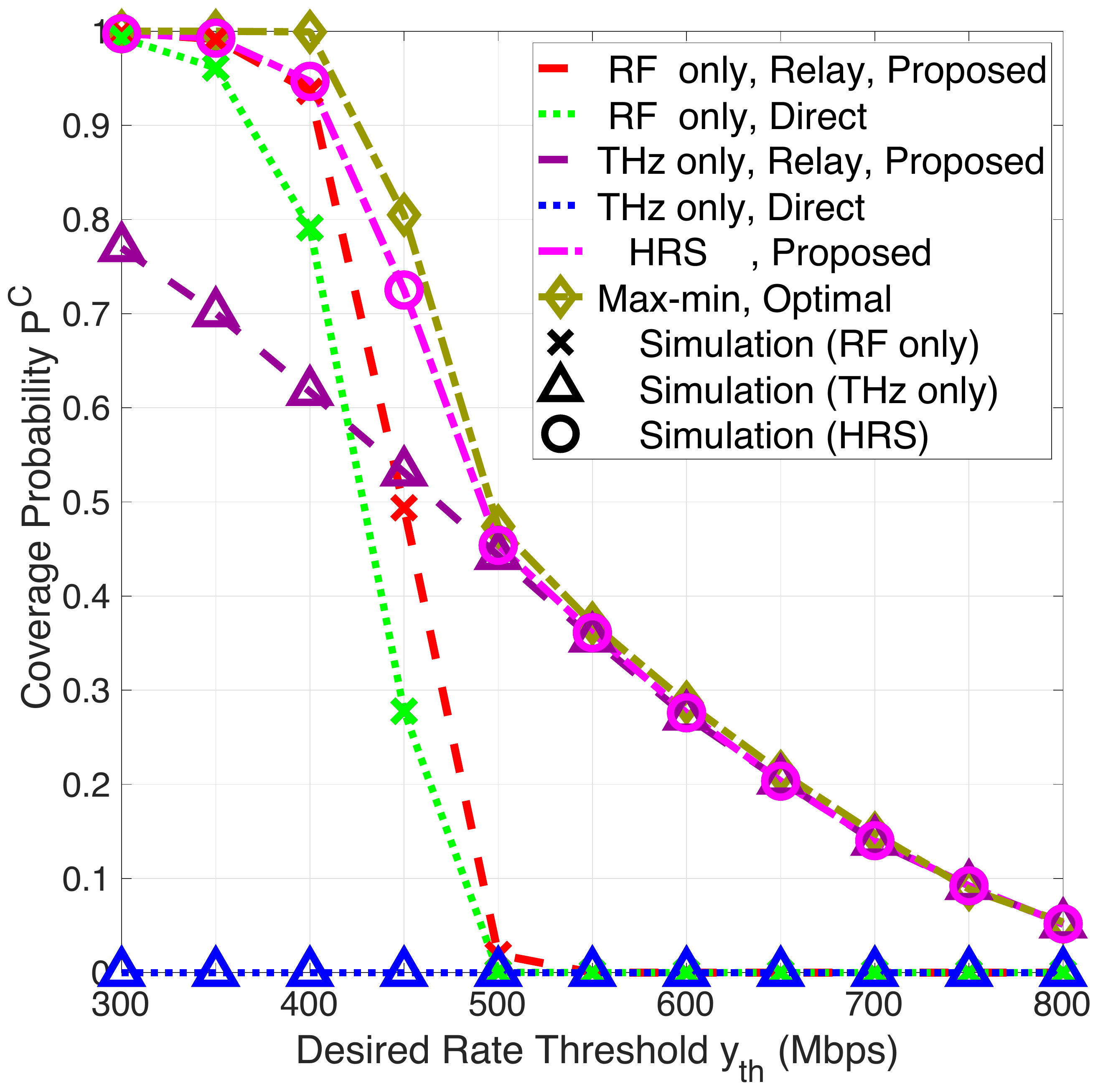}} 
    \hspace{0.4cm}
    \subfigure[$R_{SD} = 80$~m]{\includegraphics[width=0.27\textwidth]{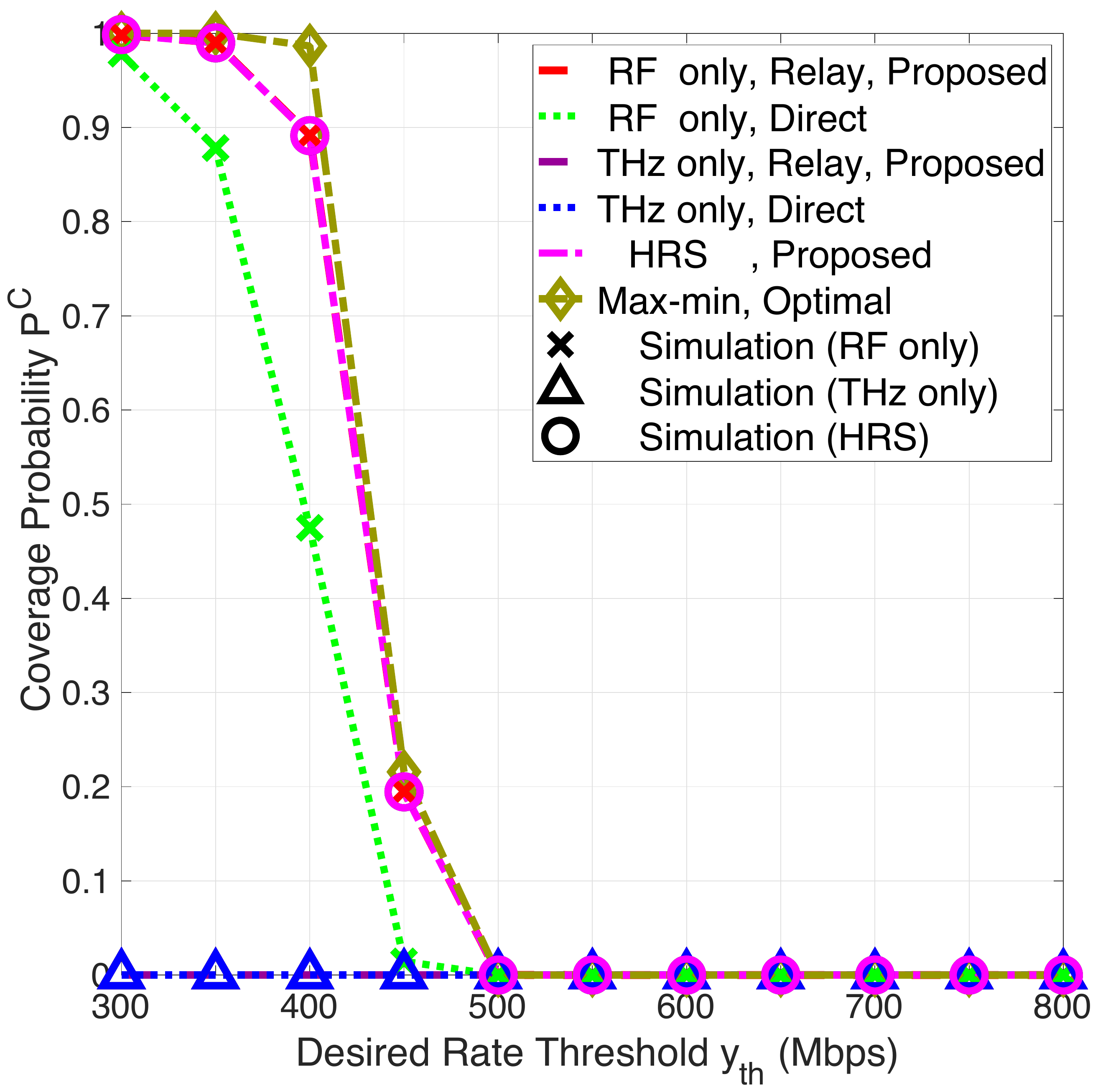}}
    \vspace{-0.3cm}
    \caption{Impact of different transmissions, the source-destination distance ($R_{SD}$), and desired data rate on coverage probability.}
    \label{fig7}
    \vspace{-0.6cm}
\end{figure*}

{\color{black}  In Fig.~\ref{fig2}, we keep $R_{SD} = 50$~m and select a set of densities with coverage of 90\% as shown in Fig.~\ref{fig1}, i.e., $\lambda_{\rm{RF}} = 5\times10^{-4}$~m$^{-2}$ and  $\lambda_{\rm{THz}} = 4\times10^{-3}$~m$^{-2}$ in HRS protocol. As shown in Fig.~\ref{fig2}, the coverage probability drops dramatically as the desired rate threshold $y_{th}$ increases from $400$~Mbps to $800$~Mbps. Moreover, comparing the blue and purple curves reveals that raising the transmission power of RF relays only improves performance when the data rate is below $500$~Mbps, but has essentially no effect when it surpasses $500$~Mbps. In general, increasing the transmission power of THz relays has a greater impact on performance than raising that of RF relays, particularly when the data rate is high.}

In Fig.~\ref{fig3}, we set $\varepsilon_{{Q}}=$~1~W and select three sets of densities with 70\% coverage as shown in Fig.~\ref{fig1}, i.e., $\lambda_{\rm{RF}}~=~\{0.45,\:1.28, \:2.27\}\times10^{-4}$~m$^{-2}$,   and  $\lambda_{\rm{THz}}~=~\{5,\;3,\;1\}\times10^{-3}$~m$^{-2}$, respectively. 
As is shown in Fig.~\ref{fig3}, the HRS coverage probability of THz relay dominated-network ($\lambda_{\rm{THz}} = 5 \times 10^{-3}$~m$^{-2}$) is strongly influenced by the distance between source and destination $R_{SD}$, especially when $40\rm{m} \leq R_{SD} \leq 60\rm{m}$. The coverage probability of THz relay-dominated network decreases drastically as $R_{SD}$ increases compared to RF relay-dominated network.

In Fig.~\ref{fig7}, we compare the proposed HRS protocol with the optimal relay selection strategy according to the \textit{max-min} selection scheme. We also compare the proposed protocol with RF-only relay selection and THz-only relay selection. {\color{black}  For the sake of fairness, we compare the HRS protocol with RF and THz direct transmissions, which use the same channel model as the relay link. Notably, direct transmission takes only one time slot, whereas relay transmission requires two \cite{belbase2018coverage}. } 
{\color{black} We set $\varepsilon_{{Q}}= 1$~W and the density of RF and THz relays  are $\lambda_{\rm{RF}} = 5\times10^{-4}$~m$^{-2}$ and  $\lambda_{\rm{THz}} = 4\times10^{-3}$~m$^{-2}$.} 
We can notice in Fig.~\ref{fig7}(a) that for short distances ($R_{SD}=20$~m), RF relay selection and RF direct transmission have lower coverage probability and it starts decreasing drastically for higher data rates ($500$~Mbps), while $P^C \approx 1$ for other four transmission schemes. Hence, we conclude from Fig.~\ref{fig7}(a) that for short distances: (1) it is better to use THz direct transmission when available than relay transmission; (2) the HRS protocol offers the same performance as the optimal protocol without requiring full and perfect CSI.
In Fig.\ref{fig7}(b), we can notice that for $R_{SD}=50$~m, the THz direct transmission has zero coverage probability. 
{\color{black} We can also notice that for lower data rates ($300-450$~Mbps), RF relay and RF direct transmission offer better performance than THz relay. In contrast, for higher data rates ($500$~Mbps),  the HRS protocol will select only THz relays since THz relay offers better performance since RF relay and RF direct transmission have a zero coverage probability.}
Finally, we notice that the proposed HRS protocol and the optimal protocol have nearly the same performance; and they both offer better performance than all the aforementioned schemes.
In Fig.\ref{fig7}(c), we can see that for large distances $R_{SD}=80$~m, THz communications (direct and relay) have $P^C=0$, and the HRS protocol will select only RF relays.


\vspace{-5mm}
\section{Conclusion}
\vspace{-1mm}


{\color{black} In this letter, the coverage probability expression of a proposed hybrid RF and THz relay selection protocol had been derived. The HRS protocol had mainly selected THz relays (associated with THz relays) for higher data rates or short source-destination distances, while it had mainly selected RF relays for lower data rates or large source-destination distances. The superiority of the HRS protocol had been demonstrated by comparing it with different strategies. In addition, the proposed HRS protocol had been compared with the optimal selection protocol, which requires a full and perfect instantaneous CSI of all the nodes in the networks. The HRS protocol had offered nearly the same performance as the optimal selection protocol for lower data rates, and the same performance for higher data rates.}

\par

\vspace{-0.3cm}
\bibliographystyle{IEEEtran}
\bibliography{references}

\end{document}